\documentclass[prd,twocolumn,showpacs,showkeys,floatfix]{revtex4}
\usepackage{amsmath,amssymb}
\usepackage{latexsym}
\usepackage{bm}
\usepackage{hyperref}

\makeatletter
\def\eqnarray{\stepcounter{equation}\let\@currentlabel=\theequation
\global\@eqnswtrue
\global\@eqcnt\z@\tabskip\@centering\let\\=\@eqncr
$$\halign to \displaywidth\bgroup\@eqnsel\hskip\@centering
  $\displaystyle\tabskip\z@{##}$&\global\@eqcnt\@ne
  \hfil${\;##\;}$\hfil
  &\global\@eqcnt\tw@ $\displaystyle\tabskip\z@{##}$\hfil
   \tabskip\@centering&\llap{##}\tabskip\z@\cr}
\makeatother

\newcommand{\smfrac}[2]{{\textstyle{#1\over#2}}}
\newcommand{\half}{\smfrac{1}{2}}
\newcommand{\df}{\text{\textrm d}}
\newcommand{\arcsinh}{\mathop{\textrm{arcsinh}}\nolimits}

\begin{document}
\title{SHEARFREE CYLINDRICAL GRAVITATIONAL COLLAPSE}
\author{A. Di Prisco}
\email{adiprisc@fisica.ciens.ucv.ve}
\author{L. Herrera}
\email{laherrera@cantv.net.ve}
\affiliation{Escuela de F\'{\i}sica, Facultad de Ciencias,
Universidad Central de Venezuela, Caracas, Venezuela.}
\author{M.A.H. MacCallum}
\email{M.A.H.MacCallum@qmul.ac.uk}
\affiliation{School of Mathematical Sciences, Queen Mary
University of London, London E1 4NS, UK.}
\author{N.O. Santos}
\email{N.O.Santos@qmul.ac.uk}
\affiliation{School of Mathematical Sciences, Queen Mary
University of London, London E1 4NS, UK. and\\
Laborat\'orio Nacional de Computa\c{c}\~ao Cient\'{\i}fica,
25651-070 Petr\'opolis RJ, Brazil.}

\begin{abstract}
We consider diagonal cylindrically symmetric metrics, with an interior
representing a general non-rotating fluid with anisotropic
pressures. An exterior vacuum Einstein-Rosen spacetime is matched to
this using Darmois matching conditions. We show that the matching
conditions can be explicitly solved for the boundary values of metric
components and their derivatives, either for the interior or
exterior. Specializing to shearfree interiors, a static exterior can
only be matched to a static interior, and the evolution in the
non-static case is found to be given in general by an elliptic
function of time. For a collapsing shearfree isotropic fluid, only a
Robertson-Walker dust interior is possible, and we show that all such
cases were included in Cocke's discussion. For these metrics, Nolan
and Nolan have shown that the matching breaks down before collapse is
complete and Tod and Mena have shown that the spacetime is not
asymptotically flat in the sense of Berger et al. The issues about
energy that then arise are revisited and it is shown that the exterior
is not in an intrinsic gravitational or superenergy radiative state at
the boundary.
\end{abstract}

\date{\today}
\maketitle


\section{Introduction and summary}

Many papers have considered cylindrical solutions, with or without
matching: see e.g.\ \cite{Bondi,Bonnor,Herrera1,Konkowski} and
Stephani et al.\ \cite{SKMHH}, Chapter 22. Here we take diagonal
metrics in both the
interior and exterior, and initially allow an anisotropic fluid
interior. The exterior is a vacuum Einstein-Rosen (ER) solution
\citep{Einstein}. Our aim was to develop a solution where one could
explicitly see the relation between source motion and gravitational
radiation, albeit in a physically unrealistic case. We were therefore
interested in collapse, though for non-static cases one can easily
reverse the sense of time so that collapse becomes expansion and vice
versa.

We set out the metrics and the Darmois matching conditions (which
preclude surface shells in the boundary) for a timelike boundary in
sections \ref{basic} and \ref{match}, and show that the junction
conditions can be explictly solved for the boundary values for the
exterior in terms of interior quantities and vice versa. This extends
the work of \cite{Herrera}.

Specializing to the shearfree case in section \ref{shearfree}, we are
able to give a first-order ordinary differential equation for the time
evolution of the interior whose solution is in general an elliptic
function. It is shown that a static exterior implies a static
interior.

Further specializing to a isotropic fluid in section \ref{PFcase}, we
prove that only a Robertson-Walker (RW) dust interior is possible and
that all such interiors are included in the discussion of Cocke
\cite{Cocke}. The matching then leads in section \ref{RWER} to
specific behaviour of the ER functions at the boundary. However,
previous work of Nolan and Nolan \cite{NolNol04} shows that the
matching breaks down before collapse is complete and Tod and Mena
\cite{TodMen04} showed that the solutions cannot be asymptotically
flat in the sense of Berger et al.\ \cite{BerChrMon95}, which makes
them unsatisfactory for our purposes.

Finally in section \ref{energy} we consider whether there are waves in
the exterior by asking if there is energy transport. Because a
cylindrically symmetric spacetime cannot be asymptotically flat, we
cannot employ the usual global definition of radiation for isolated
bodies due to \cite{BMV}, and various alternatives are discussed.
We conclude that the exterior of a cylindrical region of a collapsing
RW dust solution cannot be in an intrinsic gravitational or
superenergy radiative state at the boundary, and infer that no
radiation is transferred to or from the interior.

The specializations made here did not lead to solutions of the type we
hoped for. This is a consequence of the additional restrictions
imposed in the hope of avoiding the full complexity of the problem in
the general anisotropic case. Nevertheless we believe that it is
necessary to have the results obtained in these more restricted cases
as a first step: we hope in the future to undertake further study of
the shearing dust and shearfree anisotropic fluid interiors, which may
establish whether such results as the breakdown of the matching in the
FRW case hold more generally.

\section{Collapsing anisotropic fluid cylinders}\label{basic}

We consider a collapsing cylinder filled with anisotropic non
dissipative fluid bounded by a timelike cylindrical surface $\Sigma$ and
with energy momentum tensor given by
\begin{eqnarray}
T_{\alpha\beta}^-&=&(\mu+P_r)V_{\alpha}V_{\beta}
 +P_rg_{\alpha\beta}\nonumber \\
&& +(P_z-P_r)S_{\alpha}S_{\beta}
 +(P_{\phi}-P_r)K_{\alpha}K_{\beta},
\label{1}
\end{eqnarray}
where $\mu$ is the energy density, $P_r$, $P_z$ and $P_{\phi}$ are the
principal stresses and $V_{\alpha}$, $S_{\alpha}$ and $K_{\alpha}$ are
vectors satisfying
\begin{eqnarray}
V^{\alpha}V_{\alpha}&=&-1, \;\;
S^{\alpha}S_{\alpha}=K^{\alpha}K_{\alpha}=1,
 \nonumber\\
V^{\alpha}S_{\alpha}&=&V^{\alpha}K_{\alpha}=S^{\alpha}K_{\alpha}=0. \label{2}
\end{eqnarray}
We assume the general time dependent diagonal non rotating
cylindrically symmetric metric
\begin{equation}
ds^2_-=-A^2(dt^2-dr^2)+B^2dz^2+C^2d\phi^2, \label{3}
\end{equation}
where $A$, $B$ and $C$ are functions of $t$ and $r$. To represent
cylindrical symmetry, we impose the following ranges on the
coordinates
\begin{equation}
-\infty\leq t\leq\infty, \;\; 0\leq r, \;\; -\infty<z<\infty, \;\;
0\leq\phi\leq 2\pi, \label{4}
\end{equation}
where we assume $C=0$ at $r=0$ which is a non-singular axis.
We number the coordinates $x^0=t$, $x^1=r$, $x^2=z$ and $x^3=\phi$
and we choose the fluid to be comoving in this coordinate system;
hence from (\ref{2}) and (\ref{3})
\begin{equation}
V_{\alpha}=-A\delta^0_{\alpha}, \;\; S_{\alpha}=B\delta^2_{\alpha},
\;\; K_{\alpha}=C\delta^3_{\alpha}. \label{5}
\end{equation}
Calculating the motion of the fluid according to its expansion
$\Theta$ and shear $\sigma_{\alpha\beta}$,
\begin{eqnarray}
\Theta&=&{V^{\alpha}}_{;\alpha}, \label{6} \\
\sigma_{\alpha\beta}&=&V_{(\alpha;\beta)}
+V_{(\alpha;\gamma}V^{\gamma}V_{\beta)}-\frac{1}{3}\;
\Theta(g_{\alpha\beta}+V_{\alpha} V_{\beta}), \label{7}
\end{eqnarray}
by using (\ref{3}) and (\ref{4}) we obtain for the expansion,
\begin{equation}
\Theta=\frac{1}{A}\left(\frac{\dot{A}}{A}+\frac{\dot{B}}{B}
          +\frac{\dot{C}}{C}\right),
\label{8}
\end{equation}
and for the non zero components of the shear,
\begin{eqnarray}
\sigma_{11}&=&\frac{A}{3}\left(2\frac{\dot{A}}{A}-\frac{\dot{B}}{B}
               -\frac{\dot{C}}{C}\right),
\label{9} \\
\sigma_{22}&=&\frac{B^2}{3A}\left(2\frac{\dot{B}}{B}-\frac{\dot{A}}{A}
               -\frac{\dot{C}}{C}\right),
\label{10} \\
\sigma_{33}&=&\frac{C^2}{3A}\left(2\frac{\dot{C}}{C}-\frac{\dot{A}}{A}
                -\frac{\dot{B}}{B}\right),
\label{11}
\end{eqnarray}
where the over dot stands for differentiation with respect to
$t$. The Einstein field equations, $G_{\alpha\beta}=\kappa
T_{\alpha\beta}$, for (\ref{1}), (\ref{3}) and (\ref{5}) have the
non zero components,
\begin{eqnarray}
G^-_{00}&=&\frac{\dot{A}}{A}\left(\frac{\dot{B}}{B}+\frac{\dot{C}}{C}\right)
           +\frac{\dot{B}}{B}\frac{\dot{C}}{C} 
          -\frac{B^{\prime\prime}}{B}-\frac{C^{\prime\prime}}{C}\nonumber \\
&& +\frac{A^{\prime}}{A}\left(\frac{B^{\prime}}{B}+\frac{C^{\prime}}{C}\right)
   -\frac{B^{\prime}}{B}\frac{C^{\prime}}{C}=\kappa\mu A^2,
   \label{13}\\
G^-_{01}&=&-\frac{\dot{B}^{\prime}}{B}-\frac{\dot{C}^{\prime}}{C}
  +\frac{\dot{A}}{A}\left(\frac{B^{\prime}}{B}+\frac{C^{\prime}}{C}\right)\\
&& +\left(\frac{\dot{B}}{B}+\frac{\dot{C}}{C}\right)\frac{A^{\prime}}{A}=0,
  \label{14}\\
G^-_{11}&=&-\frac{\ddot{B}}{B}-\frac{\ddot{C}}{C}
  +\frac{\dot{A}}{A}\left(\frac{\dot{B}}{B}+\frac{\dot{C}}{C}\right)
  -\frac{\dot{B}}{B}\frac{\dot{C}}{C}
  \nonumber\\
&&+\frac{A^{\prime}}{A}\left(\frac{B^{\prime}}{B}
  +\frac{C^{\prime}}{C}\right)+\frac{B^{\prime}}{B}\frac{C^{\prime}}{C}=\kappa
  P_rA^2, \label{15}\\
G^-_{22}&=&\left(\frac{B}{A}\right)^2\left[-\frac{\ddot{A}}{A}
  -\frac{\ddot{C}}{C}+\left(\frac{\dot{A}}{A}\right)^2
  +\frac{A^{\prime\prime}}{A}
  +\frac{C^{\prime\prime}}{C}\right.\nonumber \\
&&\phantom{\left(\frac{B}{A}\right)^2[}\left.
  -\left(\frac{A^{\prime}}{A}\right)^2\right]   =\kappa P_zB^2, \label{16}\\
G^-_{33}&=&\left(\frac{C}{A}\right)^2\left[-\frac{\ddot{A}}{A}
  -\frac{\ddot{B}}{B}+\left(\frac{\dot{A}}{A}\right)^2
  +\frac{A^{\prime\prime}}{A}+\frac{B^{\prime\prime}}{B}\right.\nonumber\\
&&\phantom{\left(\frac{B}{A}\right)^2[}\left.
  -\left(\frac{A^{\prime}}{A}\right)^2\right]=\kappa P_{\phi}C^2,
  \label{17}
\end{eqnarray}
where the prime stands for differentiation with respect to $r$.

{}From (\ref{1}), (\ref{2}) and (\ref{5}) we have the non trivial
components of the Bianchi identities ${T^{\alpha}}_{\beta
;\alpha}=0$,
\begin{eqnarray}
\hspace*{-1em}
\dot{\mu}+(\mu +P_r)\frac{\dot{A}}{A}
 +(\mu+P_z)\frac{\dot{B}}{B}+(\mu+P_{\phi})\frac{\dot{C}}{C}=0,
\label{17a}\\
\hspace*{-1em}
P^{\prime}_r+(\mu+P_r)\frac{A^{\prime}}{A}
 +(P_r-P_z)\frac{B^{\prime}}{B}+(P_r-P_{\phi})\frac{C^{\prime}}{C}=0.
\label{17b}
\end{eqnarray}

For the exterior we take an ER spacetime \citep{Einstein}
\begin{equation}
ds_+^2=-e^{2(\gamma-\psi)}(dT^2-dR^2)+e^{2\psi}dz^2+e^{-2\psi}R^2d\phi^2,
 \label{m1}
\end{equation}
where $\gamma$ and $\psi$ are functions of $T$ and $R$, and for the vacuum
field equations we have
\begin{equation}
\psi_{,TT}-\psi_{,RR}-\frac{\psi_{,R}}{R}=0, \label{m2}
\end{equation}
and
\begin{equation}
\gamma_{,T}=2R\psi_{,T}\psi_{,R}, \;\;
\gamma_{,R}=R(\psi^2_{,T}+\psi^2_{,R}). \label{m3}
\end{equation}
Equation (\ref{m2}) is the cylindrically symmetric wave equation in an
Euclidean spacetime, suggesting the presence of a gravitational wave
field.

\section{Junction conditions}\label{match}

As the boundary must be comoving with the fluid interior, it will be
timelike, and given by $r=$constant in the interior metric (\ref{3})
and a curve $R(T)$ in the ER metric (\ref{m1}). Matching the collapsing
cylinder at $\Sigma$ to the ER spacetime, Darmois' junction conditions
\citep{Darmois,Herrera} give us, after a little algebra,
\begin{eqnarray}
e^{\gamma-\psi}\left[1-\left(\frac{dR}{dT}\right)^2\right]^{1/2}dT
  &\stackrel{\Sigma}{=}&Adt \stackrel{\Sigma}{=}d\tau, \label{m4}\\
e^{\psi}&\stackrel{\Sigma}{=}&B, \label{m5}\\
R&\stackrel{\Sigma}{=}&BC, \label{m6}\\
P_r&\stackrel{\Sigma}{=}&0, \label{m8}\\
e^{\psi}(R_{,\tau}\psi_{,T}+T_{,\tau}\psi_{,R})&\stackrel{\Sigma}{=}&
 \frac{B^{\prime}}{A}, \label{m7}\\
T_{,\tau}&\stackrel{\Sigma}{=}&\frac{(BC)^{\prime}}{A}.
\label{m9}
\end{eqnarray}
In this form it is easy to see that if the exterior is known, the
boundary values of $B$ and $C$, and $1/A$ times their first derivatives with
respect to $t$ and $r$ (equivalently, the
derivatives with respect to proper time in the surface and proper
distance orthogonal to it), can be solved for. The value of $A$ is not
fixed, as one can redefine the $t$ and $r$ coordinates, but its
evolution can be found from the interior field equations.

We note also that in order for the surface to be timelike we require
$(dR/dT)^2<1$.

One can reorganize these equations, together with (\ref{m3}), to
conversely give the exterior functions' values on the boundary in
terms of the interior. Rewriting (\ref{m4}), and differentiating
(\ref{m5}) and (\ref{m6}) with respect to $\tau$ in the surface, we
obtain
\begin{eqnarray}
e^{2\gamma-2\psi}(T^2_{,\tau}-R^2_{,\tau})&\stackrel{\Sigma}{=}&1,
\label{A1} \\
e^{\psi}(T_{,\tau}\psi_{,T}+R_{,\tau}\psi_{,R})&\stackrel{\Sigma}{=}
 &\frac{\dot B}{A}, \label{A2}\\
R_{,\tau}&\stackrel{\Sigma}{=}&\frac{(BC)^{\dot{}}}{A}. \label{A3}
\end{eqnarray}
Solving (\ref{A2}) and (\ref{m7}) for $\psi_{,T}$ and $\psi_{,R}$, and
substituting in (\ref{A1}), the results are (\ref{m5}) and
\begin{eqnarray}
\psi_{,T} &\stackrel{\Sigma}{=}&  \frac{B_{,t}(BC)_{,r}-B_{,r}(BC)_{,t}}
           {B[(BC)_{,r}^2 - (BC)_{,t}^2]},\label{mb1}\\
\psi_{,R} &\stackrel{\Sigma}{=}&   \frac{B_{,r}(BC)_{,r}-B_{,t}(BC)_{,t}}
           {B[(BC)_{,r}^2 - (BC)_{,t}^2]},\label{mb2}\\
e^\gamma &\stackrel{\Sigma}{=}&   \frac{AB}
           {[(BC)_{,r}^2 - (BC)_{,t}^2]^{1/2}},\label{mc1}
\end{eqnarray}
\begin{widetext}
\begin{eqnarray}
\gamma_{,T} &\stackrel{\Sigma}{=}&
           \frac{2C[B_{,t}(BC)_{,r}-B_{,r}(BC)_{,t}]
                   [(BC)_{,r}B_{,r}-(BC)_{,t}B_{,t}]}
           {B[(BC)_{,r}^2 - (BC)_{,t}^2]^2},\label{mc2}\\
\gamma_{,R} &\stackrel{\Sigma}{=}& \frac{C\{[B_{,t}(BC)_{,r}-B_{,r}(BC)_{,t}]^2+
                   [(BC)_{,r}B_{,r}-(BC)_{,t}B_{,t}]^2\}}
           {B[(BC)_{,r}^2 - (BC)_{,t}^2]^2}.\label{mc3}
\end{eqnarray}
\end{widetext}

The radius $\mathcal R$ of the collapsing cylinder as measured by the
circumference in the exterior ER spacetime is given by
\begin{equation}
{\mathcal R}\stackrel{\Sigma}{=}e^{-\psi}R\stackrel{\Sigma}{=}C. \label{m10}
\end{equation}

\section{Shearfree collapsing solution}\label{shearfree}

Now let the motion of the collapsing cylindrical fluid be shearfree,
$\sigma_{\alpha\beta}=0$. Then we can integrate (\ref{9}-\ref{11}) and
obtain
\begin{equation}
B=b(r)A, \;\; C=c(r)A, \label{18}
\end{equation}
where $b(r)$ and $c(r)$ are arbitrary functions of $r$ and the metric
(\ref{3}) becomes
\begin{equation}
ds_-^2=A^2(-dt^2+dr^2+b^2dz^2+c^2d\phi^2). \label{19}
\end{equation}
We observe that by contrast the shearfree condition in a collapsing
spherical distribution of matter in a comoving frame leaves two unknown
functions of time and radial coordinate in the metric \citep{Glass}.

Substituting (\ref{19}) into (\ref{14}) we obtain
\begin{equation}
\frac{\dot{A}^{\prime}}{A}-2\frac{\dot A}{A}\frac{A^{\prime}}{A}=0,
\label{20}
\end{equation}
which can be integrated producing
\begin{equation}
A=\frac{1}{w(t)+a(r)}, \label{21}
\end{equation}
where $w$ is an arbitrary function of $t$ and $a$ is an arbitrary
function of $r$. The expansion (\ref{8}) for (\ref{19}) and (\ref{21}) becomes,
\begin{equation}
\Theta=-3{\dot w}, \label{21a}
\end{equation}
which, as for the shearfree isotropic fluid spherical collapse,
depends also only on $t$ \citep{Glass}.

The field equations (\ref{13}-\ref{17}) with
(\ref{19}) and (\ref{21}) become
\begin{eqnarray}
\kappa \mu &=& 3\dot{w}^2+2(w+a)a^{\prime\prime}-3a^{\prime 2}
   +2(w+a)a^{\prime}\left(\frac{b^{\prime}}{b}
      +\frac{c^{\prime}}{c}\right)\nonumber\\
&&
  -(w+a)^2\left(\frac{b^{\prime\prime}}{b}
  +\frac{c^{\prime\prime}}{c}
  +\frac{b^{\prime}}{b}\frac{c^{\prime}}{c}\right), \label{22}\\
\kappa P_r &=& 2(w+a)\ddot{w}-3\dot{w}^2+3a^{\prime 2} \nonumber\\
&&  -2(w+a)a^{\prime}\left(\frac{b^{\prime}}{b}
  +\frac{c^{\prime}}{c}\right)
  +(w+a)^2\frac{b^{\prime}}{b}\frac{c^{\prime}}{c}, \label{23}\\
\kappa P_z &=& 2(w+a)\ddot{w}-3\dot{w}^2-2(w+a)a^{\prime\prime}
  +3a^{\prime2}\nonumber\\
&&-2(w+a)a^{\prime}\frac{c^{\prime}}{c}
  +(w+a)^2\frac{c^{\prime\prime}}{c},\label{24}\\
\kappa P_{\phi} &=&
  2(w+a)\ddot{w}-3\dot{w}^2-2(w+a)a^{\prime\prime}+3a^{\prime 2}
  \nonumber\\
 &&-2(w+a)a^{\prime}\frac{b^{\prime}}{b}
   +(w+a)^2\frac{b^{\prime\prime}}{b}. \label{25}
\end{eqnarray}

{}From the junction condition (\ref{m8}) we have that on the boundary
(\ref{23}) is given by
\begin{equation}
2\Omega{\ddot\Omega}-3{\dot\Omega}^2+c_2\Omega^2+c_1\Omega+c_0
 \stackrel{\Sigma}{=}0, \label{p1}
\end{equation}
where
\begin{equation}
\Omega\stackrel{\Sigma}{=}w+a, \;\; c_0
  \stackrel{\Sigma}{=}3a^{\prime 2}, \;\; c_1\stackrel{\Sigma}{=}
  -2a^{\prime}\left(\frac{b^{\prime}}{b}+\frac{c^{\prime}}{c}\right),
  \;\;
  c_2\stackrel{\Sigma}{=}\frac{b^{\prime}c^{\prime}}{bc}. \label{p2}
\end{equation}
We can integrate (\ref{p1}) producing
\begin{equation}
{\dot\Omega}^2=w_0\Omega^3+c_2\Omega^2
               +\frac{c_1}{2}\Omega+\frac{c_0}{3}, \label{p3}
\end{equation}
where $w_0$ is an integration constant. Thus in general the time
dependence is given by an elliptic function whose parameters are fixed
by values at the boundary.

If the exterior spacetime is static, $\psi_{,T}=0$, then the field
equations (\ref{m2}) and (\ref{m3}) reduce to the static Levi-Civita
spacetime,
\begin{equation}
e^{\psi}=\frac{a_1}{R^{a_2}}, \label{p6}
\end{equation}
and
\begin{equation}
e^{\gamma}=R^{a^2_2}, \label{p7}
\end{equation}
where $a_1$ and $a_2$ are integration constants: for a positive mass
source we need $a_2>0$ \cite{JenKuc94,WanDaSan97}. For the solution
(\ref{19}) and (\ref{21}) we have the following relations on
$\Sigma$. {}From (\ref{m5}) and (\ref{p6}) we have
\begin{equation}
\frac{a_1}{R^{a_2}}\stackrel{\Sigma}{=}\frac{b}{w+a}, \label{p8}
\end{equation}
and from (\ref{m5}) and (\ref{m6})
\begin{equation}
R\stackrel{\Sigma}{=}\frac{bc}{(w+a)^2}. \label{p9}
\end{equation}
With (\ref{p8}) and (\ref{p9}) we obtain
\begin{equation}
a_1(w+a)^{2a_2+1}\stackrel{\Sigma}{=}b^{a_2+1}c^{a_2}; \label{p10}
\end{equation}
since $w$ is a function of $t$ this relation can hold only if $w$ is
constant. Hence, for shearfree cylindrically symmetric anisotropic
fluids if the exterior spacetime is static, i.e. the Levi-Civita
spacetime, the cylindrical source must be static too.

Not all the junction conditions have been used, since no specific
model of the interior has been given (and static shearfree fluid
solutions may not all contain a suitable matching surface, for
instance they may not contain a surface where $P_r=0$). However, the
remaining junction conditions can be satisfied for anisotropic and
isotropic fluids, shells, and other suitable choices of interior,
see, for example, \cite{RaySom62,WanDaSan97,HerSanTei01,HerLeMar05}.

\section{Cylindrically collapsing isotropic fluid}\label{PFcase}

If the collapsing cylinder is filled with isotropic fluid then
$P_r=P_z=P_{\phi}$ and from (\ref{23}-\ref{25}) we have
\begin{eqnarray}
(w+a)\left[2\left(a^{\prime\prime}-a^{\prime}\frac{b^{\prime}}{b}\right)
 -(w+a)\left(\frac{c^{\prime\prime}}{c}
 -\frac{b^{\prime}}{b}\frac{c^{\prime}}{c}\right)\right]=0, \label{25a}\\
(w+a)\left[2\left(a^{\prime\prime}-a^{\prime}\frac{c^{\prime}}{c}\right)
 -(w+a)\left(\frac{b^{\prime\prime}}{b}-
 \frac{b^{\prime}}{b}\frac{c^{\prime}}{c}\right)\right]=0. \label{26a}
\end{eqnarray}
Then from (\ref{25a}) and (\ref{26a}), assuming $w$ has non-trivial time
dependence,
\begin{equation}
a^{\prime\prime}=a^{\prime}\frac{b^{\prime}}{b}
 =a^{\prime}\frac{c^{\prime}}{c}, \;\;
\frac{b^{\prime\prime}}{b}
 =\frac{c^{\prime\prime}}{c}
 =\frac{b^{\prime}}{b}\frac{c^{\prime}}{c}, \label{29a}
\end{equation}
which can be shown by direct calculation to reduce the Weyl tensor to
 $C_{\alpha\beta\gamma\delta}=0$, i.e.\ the spacetime inside the
 cylinder is conformally flat. All conformally flat perfect fluid
 solutions are known, and all are shearfree \cite{SKMHH}. If a
 conformally flat perfect fluid has a barotropic equation of state,
 then (Tr\"umper, cited in \cite{Ell71}) it is RW. We now show
 directly that this is the case, i.e.\ that the interior must be RW
 and thus conformally flat, without assuming barotropy, using
 regularity at the axis instead.

If $a^{\prime}\neq 0$,  then we must have
\[ \frac{a''}{a'}=\frac{b'}{b}=\frac{c'}{c},\]
from which it easily follows, using (\ref{29a}), that $a$, $b$ and $c$ are all
proportional to $e^{\beta r}$ for some non-zero constant $\beta$. This is not
consistent with having a non-singular axis where $C=0$ at $r=0$.

If $a^{\prime}=0$ then we can set $a=0$ by redefining
$w$, and from (\ref{29a}) we find that $b'/c$ and $c'/b$ are each
constant, whence
\begin{equation}
\frac{b^{\prime}}{b}\frac{c^{\prime}}{c}=-\epsilon, \label{30a}
\end{equation}
where $\epsilon$ is a constant. Writing $A=1/w$ in the field equations
(\ref{22}-\ref{25}), they become,
\begin{eqnarray}
3(A_{,\tau}^2+\epsilon)=\kappa\mu A^2, \label{31a}\\
-2AA_{,\tau\tau}-{A_{,\tau}}^2-\epsilon=\kappa PA^2. \label{32a}
\end{eqnarray}
which are easily recognizable as the usual equations for RW
spacetimes, $\epsilon$ being the spatial curvature parameter usually
denoted $k$. It is known that the most general solutions with zero
shear, rotation and acceleration are the Robertson-Walker solutions
(see e.g.\ Ellis \cite{Ell71}, page 135).  Moreover from the junction
condition (\ref{m8}), i.e.\ $P\stackrel{\Sigma}{=}0$, we have $P=0$
and consequently the fluid is a homogeneous collapsing dust, which has
no acceleration, i.e.\ a Friedman solution.

Hence we can state that \textit{a collapsing cylinder with a
non-singular axis filled with shearfree irrotational isotropic fluid
must be an RW solution and if it is matched to an ER solution the
fluid must be dust}. We again compare our result to the corresponding
isotropic spherical shearfree collapse \citep{Glass}. There the
general solution cannot be obtained since for the complete integration
of the system further equations of state are required (e.g.\ an
equation of state of the form $P=P(\mu)$ \cite{Collins}).

Now we show that the form for this discussed by Cocke \cite{Cocke} is
the most general one by direct coordinate transformations. One could
reach the final metric form more immediately by integrating
(\ref{29a}) and imposing regularity at the axis: the extra information
below is that of the coordinate transformations.

The RW metric can be expressed in spherical coordinates as
\begin{equation}
ds^2=-dt^2+A^2\left[\frac{d\bar{r}^2}{1-k\bar{r}^2}
  +\bar{r}^2(d\theta^2+\sin^2\theta\; d\phi^2)\right], \label{30}
\end{equation}
where $A$ is a function only of $t$, $k=-1$, $0$ or $1$, and the ranges of the
coordinates are
\begin{equation}
-\infty\leq t\leq\infty, \;\; 0\leq\bar{r}<\infty, \;\;
 0\leq\theta\leq\pi, \;\; 0\leq\phi\leq 2\pi, \label{31}
\end{equation}
except when $k=1$ where instead $0\leq\bar{r}<1$ and $r=1$ is the
antipode of the origin.
To write (\ref{30}) in cylindrical coordinates we make the transformation
\begin{equation}
\hat{r}=\bar{r}\sin\theta, \;\; z=f(\bar{r},\theta), \label{32}
\end{equation}
which yields
\begin{equation}
\df\hat{r}=\sin\theta\; \df\bar{r}+\bar{r}\cos\theta\; \df \theta, \;\;
 \df z=f_{,\bar{r}}\;\df\bar{r}+f_{,\theta}\;\df\theta. \label{33}
\end{equation}
The inverse transformation is
\begin{equation}
\df\bar{r}=\frac{f_{,\theta}\;\df\hat{r}-\bar{r}\cos\theta\;
  \df z}{\sin\theta\;f_{,\theta}-\bar{r}\cos\theta \;f_{,\bar{r}}}, \;\;
\df\theta=\frac{\sin\theta\;\df z-f_{,\bar{r}}\;
 \df\hat{r}}{\sin\theta\;f_{,\theta}-\bar{r}\cos\theta\;f_{,\bar{r}}}.
 \label{34}
\end{equation}
In order to write (\ref{30}) transformed by (\ref{32}) we first observe that
\begin{widetext}
\begin{equation}
\frac{\df\bar{r}^2}{1-k\bar{r}^2}+\bar{r}^2\df\theta^2=
\frac{\displaystyle{\left[\left(\frac{f^2_{,\theta}}{1-k\bar{r}^2}
 +\bar{r}^2f^2_{,\bar{r}}\right)\df\hat{r}^2 
-2\bar{r}\left(\frac{\cos\theta\;f_{,\theta}}{1-k\bar{r}^2}
+\bar{r}\sin\theta\;f_{,\bar{r}}\right)\df\hat{r}\;\df z
+\frac{\bar{r}^2(1-k\bar{r}^2\sin^2\theta)}{1-k\bar{r}^2}\;\df
z^2\right]}}{(\sin\theta\;f_{,\theta}-\bar{r}\cos\theta\;f_{,\bar{r}})^2}.
 \label{35}
\end{equation}
\end{widetext}
Imposing the requirement that the transformation does not produce
cross terms we must choose
\begin{equation}
f_{,\theta}=-\bar{r}(1-k\bar{r}^2)\frac{\sin\theta}{\cos\theta}f_{,\bar{r}}.
 \label{36}
\end{equation}
Substituting (\ref{36}) back into (\ref{35}) we obtain
\begin{equation}
\frac{\df\bar{r}^2}{1-k\bar{r}^2}+\bar{r}^2\df\theta^2
 =\frac{\displaystyle{\left[\df\hat{r}^2
       +\frac{\cos^2\theta}{(1-k\bar{r}^2)f_{,\bar{r}}^2}\;
 \df z^2\right]}}{1-k\bar{r}^2\sin^2\theta}. \label{37}
\end{equation}
With (\ref{32}) and (\ref{37}) we can write (\ref{30}) as
\begin{equation}
ds^2=-dt^2+A^2\left(\frac{d\hat{r}^2}{1-k \hat{r}^2}+h^2dz^2
 +\hat{r}^2d\phi^2\right), \label{38}
\end{equation}
where
\begin{equation}
h=\frac{\cos\theta}{(1-k\bar{r}^2\sin^2\theta)^{1/2}
 (1-k\bar{r}^2)^{1/2}f_{,\bar{r}}}. \label{39}
\end{equation}
Then from (\ref{36}) and (\ref{39}) we have
\begin{eqnarray}
f_{,\bar{r}}&=&\frac{\cos\theta}{h(1-k\bar{r}^2\sin^2\theta)^{1/2}
  (1-k\bar{r}^2)^{1/2}}, \label{40}\\
f_{,\theta}&=&-\frac{\bar{r}\sin\theta(1-k\bar{r}^2)^{1/2}}
 {h(1-k\bar{r}^2\sin^2\theta)^{1/2}}. \label{41}
\end{eqnarray}
Requiring that $h$ is a function only of $\hat{r}$ we have
\begin{equation}
h_{,\bar{r}}=\sin\theta\frac{\df h}{\df\hat{r}}, \;\;
h_{,\theta}=\bar{r}\cos\theta\frac{\df h}{\df\hat{r}}. \label{42}
\end{equation}
Differentiating (\ref{40}) and (\ref{41}) with respect to $\theta$ and
$\bar{r}$ respectively and using (\ref{42}) we obtain
\begin{widetext}
\begin{eqnarray}
\hspace*{-2em}
f_{,\bar{r}\theta}&=&-\frac{1}{h(1-k\bar{r}^2\sin^2\theta)^{1/2}
  (1-k\bar{r}^2)^{1/2}} 
\left[\frac{(1-k\bar{r}^2)\sin\theta}
  {1-k\bar{r}^2\sin^2\theta}+\frac{\bar{r}\cos^2\theta}{h}\frac{dh}{d\hat{r}}
  \right], \label{43}\\
\hspace*{-2em}
f_{,\theta\bar{r}}&=&-\frac{\sin\theta}{h(1-k\bar{r}^2\sin^2\theta)^{1/2}}
  \left[\frac{1-2k\bar{r}^2}
  {(1-k\bar{r}^2\sin^2\theta)^{1/2}}
+\frac{k\bar{r}^2\sin^2\theta(1-k\bar{r}^2)^{1/2}}
  {1-k\bar{r}^2\sin^2\theta}-
  \frac{\bar{r}\sin\theta(1-k\bar{r}^2)^{1/2}}{h}\frac{dh}{d\hat{r}}\right].
  \label{44}
\end{eqnarray}
\end{widetext}
Equating the two expressions (\ref{43}) and (\ref{44}), i.e.\ imposing
the integrability condition
$f_{,\bar{r}\theta}=f_{,\theta\bar{r}}$, and using (\ref{32})
we obtain
\begin{equation}
\frac{1}{h}\frac{\df h}{\df\hat{r}}=-\frac{k \hat{r}}{1-k \hat{r}^2},
 \label{45}
\end{equation}
and integrating this we obtain
\begin{equation}
h=\gamma(1-k \hat{r}^2)^{1/2}, \label{46}
\end{equation}
where $\gamma$ is an integration constant. Now substituting (\ref{46})
into (\ref{38}) and rescaling $z$ we finally have
\begin{equation}
ds^2=-dt^2+A^2\left[\frac{d\hat{r}^2}{1-k \hat{r}^2}+(1-k
\hat{r}^2)dz^2+\hat{r}^2d\phi^2\right]. \label{47}
\end{equation}
The metric form (\ref{47}) was obtained by Cocke \cite{Cocke}, but here we
have proved that it is the general RW metric in cylindrical
coordinates.

In order to write (\ref{47}) in the coordinate system employed in
(\ref{3}) we rescale $t$ and transform $\hat{r}$ so that
\begin{equation}
r=\int\frac{d\hat{r}}{(1-k\hat{r}^2)^{1/2}}, \label{48}
\end{equation}
and we obtain
\begin{equation}
ds^2=A^2(t)(-dt^2+dr^2+g^{\prime 2}\;dz^2+g^2\;d\phi^2), \label{49}
\end{equation} or
\begin{equation}
ds^2=-d\tau^2+A^2(\tau)(dr^2+g^{\prime 2}\;dz^2+g^2\;d\phi^2), \label{49a}
\end{equation}
with $g=\sinh r$, $r$ or $\sin r$ according as $k=-1$, $0$ or $1$.

The metric (\ref{19}), with (\ref{21}) and (\ref{30a}), is the general
RW metric (\ref{49}) where $k=\epsilon$ and $w=1/A$, and it
satisfies (\ref{31a}) and (\ref{32a}).

\section{Matching FRW spacetime to ER spacetime}\label{RWER}

Matchings between a cylindrical homogeneous perfect fluid interior and
a vacuum exterior (or vice versa) have been studied by Mena et al.\
\cite{MenTavVer02}, who showed that matching to a static vacuum is
impossible which is a special case of our result above, by Nolan and
Nolan \cite{NolNol04}\footnote{We are grateful to the anonymous
referee for drawing the works of Hayward and of Nolan and Nolan to our
attention, for picking out points which suggested to us how to
integrate the latter with our discussion, and for a considerable
number of other pertinent comments.}, and by Tod and Mena
\cite{TodMen04}. In the last of these papers, the matching of ER and
$k=0$ FRW metrics is given, using coordinates in which the exterior
metric takes the form
\begin{equation}
ds_+^2=-e^{2(\gamma-\psi)}(d\hat{T}^2-d\rho^2)+e^{2\psi}dz^2
 +e^{-2\psi}R^2d\phi^2,  \label{m1a}
\end{equation}
where $R=R(\hat{T},\,\rho)$: the main difference from our treatment is
that the coordinates $(\hat{T},\,\rho)$ are chosen so that the
boundary is at $\rho=$ constant. Here $\hat{T}$ has replaced the $T$
of the original paper to avoid confusion. Tod and Mena \cite{TodMen04}
study the global structure and conclude that the spacetime is not
asymptotically flat in the sense of Berger et al.\ \cite{BerChrMon95},
but instead has a singular Cauchy horizon.

In \cite{NolNol04}, the matching of a general cylindrically symetric
vacuum to FRW is studied: the exterior is then shown, as one might
expect, to be of ER form. Here we shall find the trajectory of the
boundary as an equation relating $T$ and $R$, and then display the
conditions satisfied there by $\gamma$ and $\psi$. Then we comment
further on the results of Tod and Mena \cite{TodMen04} and Nolan and
Nolan \cite{NolNol04}.

The well-known solutions to (\ref{31a}) and (\ref{32a}) are
\begin{eqnarray}
k&=1,  &~~A=m\,\sin^2\Psi/2, \quad 2\tau=m(\Psi-\sin\Psi),
\label{eq:12.3a} \\
k&=0,  &~~A=(3\sqrt{m}\tau/2)^{2/3}, \label{eq:12.3b} \\
k&=-1, &~~A=m\sinh^2\Psi/2,
\quad 2\tau=m(\sinh\Psi- \Psi), \label{eq:12.3c}
\end{eqnarray}
where $m$ is a constant giving the mass density, see e.g.\
\cite{SKMHH}, equations (14.6). These are given in a form expanding
as $\tau$ increases from 0. Reversing the sense of $\tau$ in
(\ref{eq:12.3a}--\ref{eq:12.3c}), so that we have
collapse, and introducing a
constant $T_0$ where the singularity occurs, we use these in $R=A^2
gg'$ and $T_{,\tau} = A(gg')'$ (which follow from (\ref{m6}),
(\ref{m9}) and (\ref{49a})). We can then integrate for $T$, and where
necessary eliminate $\Psi$, to get
\begin{eqnarray}
2\alpha_-(T_0-T)&\stackrel{\Sigma}{=}&
 R^{1/4}\left(R^{1/2}-\frac{3}{2}R_0\right)(R^{1/2}+R_0)^{1/2} \nonumber\\
&&\phantom{junk}
 +\frac{3}{2}R_0^2\arcsinh\left(\frac{R^{1/4}}{R_0^{1/2}}\right),
 \label{59}\\
T_0-T&\stackrel{\Sigma}{=}&R_0R^{5/4}, \label{60}\\
2\alpha_+(T_0-T)&\stackrel{\Sigma}{=}&
 -R^{1/4}\left(R^{1/2}+\frac{3}{2}R_0\right)(R_0-R^{1/2})^{1/2} \nonumber\\
&&\phantom{junk}
  +\frac{3}{2}R_0^2\arcsin\left(\frac{R^{1/4}}{R_0^{1/2}}\right), \label{61}
\end{eqnarray}
where $T_0$ and $R_0$ are integration constants, and
\begin{equation}
\alpha\stackrel{\Sigma}{=}
 \frac{gg^{\prime}}{(gg^{\prime})^{\prime}}, \label{56}
\end{equation}
evaluated on $\Sigma$,
and thus from (\ref{49}) for $\epsilon=-1$ and $1$ respectively
\begin{eqnarray}
\alpha_-&\stackrel{\Sigma}{=}&\frac{\tanh r_0}{1+\tanh^2 r_0}, \label{57}\\
\alpha_+&\stackrel{\Sigma}{=}&\frac{\tan r_0}{1-\tan^2 r_0}, \label{58}
\end{eqnarray}
where $r\stackrel{\Sigma}{=}r_0$, while for $k=0$, $\alpha=r_0$.

On this moving boundary we know from (\ref{m5}) and
(\ref{mb1}-\ref{mc1}) that we must have (noting that for all three
possible $g$, $g'^2-gg'' = 1$)
\begin{eqnarray}
e^\psi &\stackrel{\Sigma}{=}& Ag',\\
\psi_{,T}&\stackrel{\Sigma}{=}&\frac{A_{,\tau}}
                 {A^2((gg')')^2[1-(2\alpha A_{,\tau})^2]},\\
\psi_{,R}&\stackrel{\Sigma}{=}&\frac{-g(k(gg')' +2A_{,\tau}^2g'^2)}
                 {A^2((gg')')^2[1-(2\alpha A_{,\tau})^2]},\\
e^\gamma&\stackrel{\Sigma}{=}& \frac{g'}
                 {(gg')'\sqrt{1-(2\alpha A_{,\tau})^2}},
\end{eqnarray}
where the functions of $r$ have to be evaluated at $r_0$.

One can give a general solution of (\ref{m2}) as a sum of separable
solutions but we have not been able to determine the specific solution
which matches to RW even in the simplest ($k=0$) case.

Our condition that $R$ be a spatial coordinate in the exterior metric
is violated when $T$ is sufficiently close to $T_0$: in fact as $T$
increases
\[ \left| \frac{\df R}{\df T} \right| \rightarrow 1 \Leftrightarrow
5R_0^{4/5}(T_0-T)^{1/5}\rightarrow 4.\] (Note that this can also be
expressed as $1=|2\alpha A_{,\tau}|$.) For larger $T$ our coordinates,
and hence our matching, do not apply at $\Sigma$. This agrees with the
results of Nolan and Nolan
\cite{NolNol04}, who showed that such a breakdown is inevitable, essentially
because the collapsing source always leads to trapped cylinders,
whereas the ER vacuum region cannot contain trapped cylinders
\cite{Thorne}. They describe the matching as impossible, meaning that
it cannot be carried right up to the FRW singularity, but note that it
can be used up to some finite time.

One can continue the discussion of trapped cylinders by using the
coordinates of Tod and Mena \cite{TodMen04}.  The bound on
applicability of our matching corresponds to their conclusion that
in the $k=0$ FRW case the boundary becomes a marginally trapped
surface when $1=|2\alpha A_{,\tau}|$ (in their notation, with hats added
for clarity, at $\hat{T}=\hat{\alpha}- (4\hat{r}_0/3)^3$) so at larger
$T$ the surface is trapped.  Such a trapped surface is not consistent
with ``asymptotic flatness'' in the sense of Berger et al.\
\cite{BerChrMon95} (see their Proposition 2.3).

Tod and Mena further show that $u=\sqrt{R}\psi$ (in their coordinates)
has a divergent derivative at the Cauchy horizon (essentially the past
null cone of the FRW singularity) and hence conclude that this horizon
is singular. The argument for these conclusions applies also to the
$k=\pm 1$ cases, with changes in formulae. Tod and Mena infer that
there is incoming gravitational radiation: however, this does not seem
to be supported by the calculations in the following section.

\section{Energy, superenergy, radiation and boundary conditions}\label{energy}

Part of our motivation was a search for an exact cylindrical solution
for interior and exterior enabling one to study exactly how
gravitational radiation arises. Our ans\"atze for the interior turned out
to allow only FRW, which we would expect to be
non-radiative in any definition. Thus we expect the exterior to also be
non-radiative (or possibly to show waves coming in from infinity and
totally reflected at the boundary with the interior).

A cylindrically symmetric spacetime cannot be asymptotically flat, due
to the behaviour in directions parallel to the axis, so one cannot
employ the usual global definition of radiation for isolated bodies
due to Bondi et al.\ \cite{BMV}. Moreover, it appears from Proposition
2.3 of Berger et al.\ \cite{BerChrMon95} that simple cylindrical
solutions with collapsing cores could not even be ``asymptotically
flat'' in their modified sense, because one would expect trapped
cylindrical surfaces to arise in any collapse which does not halt or
reverse, and such surfaces are not compatible with ``asymptotic
flatness''.

So for more detailed study we need some definition of radiation, or
energy, or energy density, other than the one from asymptotic
flatness. This must be (quasi-)local, at least in $z$, to avoid the
problem that integration for $z$ from $-\infty$ to $\infty$ would
obviously give an infinite answer for any non-zero energy density.

There is also the possibility of energy being transferred to or from
any given region by transport along the axis. Bondi \cite{Bondi} showed
that there is no conserved mass per unit length as a result of
`intangible' gravitational induction arising from axial motion (and
distinct from the work done by axial pressure).

It is well known that in relativity there cannot be any covariant
local definition of energy, as this would violate the equivalence
principle. There are a number of quasi-local definitions, using either
integrals over surfaces, or pseudo-tensors integrated over volumes
(which in general also reduce to surface integrals, provided the
interior volumes do not have discontinuities or singularities of the
pseudo-tensor). Clavering \cite{Clavering} has shown that none of the
latter agree with the various quasi-local surface integrals discussed
by Szabados \cite{Szabados}, and none of them are satisfactory, for a
variety of reasons, the best being M\"oller's definition. We therefore
do not calculate the pseudo-tensorial energies.

In his study of definitions of standing waves, Stephani
\cite{Stephani} considered cylindrical systems. His results suggest
that for the ER solutions Thorne's ``C-energy'' \cite{Thorne} is the
least unsatisfactory.  (A recent indication of the unsatisfactoriness
of C-energy, even in vacuum, has been given in \cite{HarNakNol08},
where it is shown that it can be non-vanishing in Minkowski space.)
We consider, following Chiba \cite{Chi96} and Hayward \cite{Hay00},
the modified C-energy defined in the ``Note added in proof'' on pages
B256-257 of Thorne's paper. Taking the generic cylindrically symmetric
metric in the form
\begin{equation}
ds^2=-e^{2(\gamma-\psi)}(dT^2-dr^2)+e^{2\psi}dz^2+e^{-2\psi}R^2d\phi^2,
 \label{m1mod}
\end{equation}
where $R=R(r,\,t)$, this energy is
\begin{equation}
E = \smfrac{1}{8}(1-(R^{\prime 2}-\dot{R}^2)e^{-2\gamma})
\label{Cenergy}
\end{equation}
where the prime and dot refer to differentiation with respect to $r$
and $t$ respectively.

Hayward has shown, in an elegant formulation, that one can define an
invariant tensor (in his notation, $\Theta_{ab}$) such that the sum of
this and the usual energy-momentum tensor $T_{ab}$ is conserved: $\Theta$
may then be interpreted as the energy-momentum of gravitational waves. One
can derive the conservation in a simple way from the fact that for the metric
(\ref{m1mod}), when matter is
present, the equations (\ref{m3}) generalize to
\begin{eqnarray}
\half[(R^{\prime 2}-\dot{R}^2)e^{-2\gamma}]' &=&  -RR'(\kappa
T_{00}+\psi^{\prime 2} +\dot{\psi}^2) \\
&&\phantom{zzz}+R\dot{R}(\kappa T_{01} \nonumber
+2\psi^\prime\dot{\psi}),\\ 
\half[(R^{\prime 2}-\dot{R}^2)e^{-2\gamma}]\dot{} &=&  -RR'(\kappa T_{01}
+2\psi^\prime\dot{\psi}) \\
&&\phantom{zzz}+R\dot{R}(\kappa T_{11}+\psi^{\prime 2} +\dot{\psi}^2)
\nonumber
\end{eqnarray}
(cf.\ equations (A41), (A42), of \cite{Hay00}): the result is then
just the integrability condition for the left sides, the relevant
terms of $\Theta$ being the terms in $\psi$ on the right sides. That
these terms are invariantly defined (provided there are no more
translational Killing vectors) arises because the two Killing vectors,
$\partial_\phi$ and $\partial_z$, and their lengths, are uniquely
defined up to rescaling of $z$, so $\psi$ is fixed up to a constant
and $R$ up to a constant factor.

Hayward's discussion also brings out the fact that for the form
(\ref{m1mod}), if $\mu$ and $\rho$ are the divergences of the incoming
and outgoing null normals to timelike cylinders of symmetry, then
\begin{equation}
\rho \mu = \frac{1}{8R^2} e^{2(\psi-\gamma)}(R^{\prime 2}-\dot{R}^2)
\label{murho}
\end{equation}
with an obvious relation to (\ref{Cenergy}).

This is also related to one of the best known quasi-local energy
definitions by an integral on a surface, the Hawking mass.  For a
closed surface $S$ with surface area element $\df {\textbf S}$ this
mass is
\[ m = \frac{1}{(4\pi)^{3/2}}\left(\oint_S \df{\textbf
  S}\right)^{1/2}\left(2\pi - \oint_S \mu\rho\,\df{\textbf
  S}\right). 
\] 
To make a closed cylindrical surface we would have to add (e.g.)
surfaces $z=$ constant at
some finite $z$ values, but $\mu\rho$ has the same values for all $z$
on such surfaces and so we could ignore them for very large
cylindrical surfaces and think of $E$ as giving a Hawking mass per unit
length in $z$.

Unfortunately this does not lead to a satisfactory account of energy
lost or gained by the collapsing dust. The Darmois conditions imply
that $E$ is continuous at $\Sigma$ (one can check this by direct
calculation, but it is easy to understand because there is no surface
layer and hence no immediate change in geodesic deviations at
$\Sigma$, so $\mu\rho$ is continuous, and the continuity of
$g_{\phi\phi}$ and $g_{zz}$ then shows the same is true for $E$.)
Calculating on the dust side of $\Sigma$, using (\ref{49}), we have
\[ E=\frac{1}{8}
\left(1-\frac{((gg')')^2}{g'^2} +4g^2 \frac{\dot{A}^2}{A^2}\right).
\]
(As one expects, \cite{Hay00}, this is $O(r^2)$ as $r\rightarrow 0$.)
At a fixed $r$, only the ${\dot{A}^2}/{A^2}=A_{,\tau}^2$ term changes,
and it increases as the dust collapses. This is consistent with
Cocke's calculation \cite{Cocke} of (unmodified) C-energy flux, normal
to cylinders of constant $R$, in our notation, in the Einstein-Rosen
region, where he found an inward flux.

It would, however, be physically very odd if this had to be
interpreted as the exterior giving energy to the dust, since the
behaviour of the dust is exactly the same as in a uniform universe,
where a cylinder cannot easily be thought of as taking energy from
the rest of the universe. Probably the result is better considered as
another indication of the unsatisfactoriness of C-energy.

Other local characterizations of the presence of radiation are given
by the decomposition of the Bel-Robinson tensor. The behaviour of this
tensor in an ER spacetime was considered in \cite{h}, where the
obvious unit vector in the form (\ref{m1}), i.e.\ the one parallel to
$\partial/\partial T$, was used to define the decomposition. In that
reference it was shown that for a pulse of radiation, well behind the
front of the pulse, there is an incoming flux of superenergy, which
seems to agree with Cocke's result. However, to define states of
intrinsic radiation one has to consider all possible timelike vectors,
and we now show, using the timelike vector parallel to the boundary,
that our solutions do not have intrinsic radiation there according to this
definition.

Bel \cite{Bel,Bel2} defined the tensor
\begin{eqnarray}
T^{abcd} &=& R^{aecf}{{{R_e}^b}_f}^d +{}^*R^{aecf}{}^*{{{R_e}^b}_f}^d
 \nonumber\\
&&+R^{*aecf}{{{{R^*}_e}^b}_f}^d+ {}^*R^{*aecf}{}^*{{{{R^*}_e}^b}_f}^d,
\label{belrob}
\end{eqnarray}
which in the vacuum case is referred to as the Bel-Robinson
tensor. Here the star operation is the usual Hodge dual (see e.g.\
\cite{SKMHH}, chapter 3).  \citet{BonSen97} pointed out that with the
usual decomposition in terms of the Weyl and Ricci tensors, $C_{abcd}$
and $R_{ab}$, i.e.\
\begin{equation}
\label{eq:3.44}R_{abcd}=C_{abcd}+E_{abcd}+G_{abcd},
\end{equation}
where
\begin{eqnarray}
E_{abcd} &\equiv& \half(g_{ac}S_{bd}+g_{bd}S_{ac}-g_{ad}S_{bc}-g_{bc}S_{ad}),
\label{eq:3.45}\\
G_{abcd}&\equiv
&\smfrac{1}{12}R(g_{ac}g_{bd}-g_{ad}g_{bc})\equiv\smfrac{1}{12}Rg_{abcd},
\label{eq:3.46} \\
S_{ab}&\equiv &R_{ab}-\smfrac{1}{4}Rg_{ab}, \qquad R\equiv R^{a}{}_{a},
\label{eq:3.47}
\end{eqnarray}
the Bel tensor can be written as
\begin{eqnarray}
T^{abcd} &=& C^{aecf}{{{C_e}^b}_f}^d+{}^*C^{aecf}{}^*{{{C_e}^b}_f}^d
 \nonumber\\
&&           +\frac{R}{6}(C^{acbd}+C^{adbc})+M^{abcd},
\end{eqnarray}
where the matter contribution is
\begin{eqnarray}
M_{abcd} &=&S^{ab}S^{cd}\nonumber\\
&& +\half S^{ae}{S^{b}}_e g^{cd}+ \half
S^{ce}{S^{d}}_e g^{ab}
         - 2 S^{e(a} g^{b)(c}{S^{d)}_e}\nonumber \\
&& +\smfrac{1}{4} S^{ef} S_{ef}(g^{ac} g^{bd} + g^{ad} g^{bc} - g^{ab}
	   g^{cd}) \label{belmatter}\\
&& +\frac{R^2}{144}( 2 g^{ac} g^{bd} + 2 g^{ad} g^{bc} - g^{a b}
	   g^{cd}).\nonumber
\end{eqnarray}

One can decompose the Bel tensor in a 3+1 formalism \citep{Lobo}
relative to a unit timelike vector $n^a$, giving, among the parts, the
superenergy $W$, the superPoynting vector $P^a$ and the tensor
$Q_{abc}$ defined by
\begin{eqnarray}
W &\equiv& T^{abcd} n_a n_b n_c n_d, ~~
P^d \equiv T^{abce}n_a n_b n_c {h^d}_e,\nonumber \\
Q^{bcd} &\equiv&  -T^{aefg}n_a {h^b}_e {h^c}_f {h^d}_g,
\label{sebits}
\end{eqnarray}
where $h_{ab}=g_{ab}+n_a n_b$.
\citet{Bel2} defined a state of intrinsic gravitational radiation
(at a point $p$) to be one in which $P^a \neq 0$ for any choice of
$n^a$ (at $p$). \citet{Lobo} similarly defines an
intrinsic superenergy radiative state to be one where $Q^{abc} \neq 0$
for any choice of $n^a$. In vacuum,
\begin{eqnarray}
P^a &=& 2{B_p}^l E_{ql}\eta^{apq},\nonumber \\
\quad Q_{cdb} &=& h_{cd}P_b -
2(B_{da} E_{cf} - B_{ca} E_{df}){\eta_b}^{af},\label{PQ}
\end{eqnarray}
where, as usual, $E_{ab} = C_{acbd}n^b n^d$, $B_{ab}={}^*C_{acbd}n^b
n^d$, and $\eta_{abc} = \eta_{abcd} n^d$, $\eta_{abcd}$ being the
usual volume 4-form density.

We now consider how these quantities behave at the (timelike) boundary
between regions of spacetime. It is simplest to describe this using an
orthonormal tetrad $\{{\bm e}_a,~ a=0...3\}$ chosen such that the
timelike unit vector ${\bm e}_0$ lies on the boundary surface (and
will be used as $n^a$ in (\ref{sebits})) and ${\bm e}_1$ is the
normal to the surface. From the equations (80) of
\citet{MarSen93} we can straightforwardly show that if the Darmois
junction conditions are satisfied then the following combinations of
Riemann tensor components are continuous:
\begin{eqnarray}
&& E_{12}, ~~R_{12}, ~~E_{13}, ~~R_{13},  ~~R_{01},\\
&& B_{11},
~~B_{22} ~~(\textrm{and hence}~B_{33}), ~~B_{23},~~\label{line1}\\
&& B_{13}+\half R_{02}, ~~B_{12} - \half R_{03}, ~~E_{23}-\half
  R_{23}\label{line2}\\
&& -E_{11}-\smfrac{1}{6}R +\half(R_{22}+R_{33}),
  ~~E_{22}+\smfrac{1}{6}R +\half(R_{00}-R_{22}),\nonumber\\
&& E_{33}+\smfrac{1}{6}R +\half(R_{00}-R_{33}).\label{line3}
\end{eqnarray}
Note that one consequence is (\ref{m8}), the
continuity of $T_{11}$.

{}From (\ref{line1}) and (\ref{line2}) we see that (assuming we align
${\bm e}_2$ and ${\bm e}_3$ with $S_a$ and $K_b$) then
with the energy-momentum form assumed in (\ref{1}), $B_{ab}$ is
continuous at the boundary. Moreover, the reflection symmetries in
$\phi$ and $z$ imply that $P_2=P_3 =0$ (and $B_{12} = B_{13} =0$) in
both interior and exterior (note that since reversing one axis also
reverses the orientation and hence the sign of the dual, it does not
follow that $B_{23}=0$). 

{}From this, it is easy to see that for an FRW interior, which is
conformally flat, $B_{ab}=0$ on both sides of the boundary, so from
(\ref{PQ}) $P_a=0$ and $Q_{abc}=0$ there in the frame defined above
(i.e.\ with $n^a = {\bm e}_0^a$ and under Darmois boundary
conditions). Hence at the boundary the exterior does not have
intrinsic gravitational radiation or intrinsic superenergy
radiation. Thus it is reasonable to conclude that any possible
radiation in the exterior spacetime is not produced by the
source. This agrees with our expectation that such an interior should
not radiate or absorb radiation. Note that we have not excluded the
possibility of total reflection at the boundary if at a reflecting
surface these indicators ($P^a$ and $Q_{cdb}$) would show no intrinsic
radiation, and it is therefore possible that there could be non-zero
incoming and/or outgoing radiation in the exterior. The conclusion
that no radiation crosses the boundary seems in conflict with the
discussion given by \citet{TodMen04}.

\begin{acknowledgments}
LH wishes to thank FUNDACION EMPRESAS POLAR for financial support and
the University of the Basque Country. ADP acknowledges the hospitality
of the University of the Basque Country and financial support from the
CDCH at Universidad Central de Venezuela. LH and ADP also acknowledge
financial support from the CDCH at Universidad Central de Venezuela
under grants PG 03-00-6497-2007.

NOS gratefully acknowledges financial assistance from the United
Kingdom EPSRC under grant EP/E063896/1 and from CNPq Brazil.
\end{acknowledgments}


\end{document}